\documentclass[12pt]{iopart}
\usepackage{epsfig}
\usepackage{citesort}
\usepackage{a4wide}

\newcommand{\gtrsim}{\,\rlap{\lower3.7pt\hbox{$\mathchar\sim$}}
\raise1pt\hbox{$>$}\,}
\newcommand{\lesssim}{\,\rlap{\lower3.7pt\hbox{$\mathchar\sim$}}
\raise1pt\hbox{$<$}\,}

\newcommand{\gwig}{\mbox{\;\raisebox{.3ex}
    {$>$}$\!\!\!\!\!$\raisebox{-.9ex}{$\sim$}}\;}
\def\Geff{G_{\rm eff}}

\begin{document}
{\hfill\normalsize \tt DESY 06-230}

\title{Neutrino Dark Energy -- Revisiting the Stability Issue}

\author{Ole Eggers Bj{\ae}lde$^1$, Anthony W. Brookfield$^2$, Carsten van de Bruck$^3$,
Steen Hannestad$^1$, David F. Mota$^{4,5}$, Lily Schrempp$^6$, and
Domenico Tocchini-Valentini$^7$}
\address{$^1$ Department of Physics and Astronomy, University of Aarhus,
Ny Munkegade, DK-8000 Aarhus C, Denmark}
\address{$^2$ Department of Applied Mathematics and Department of Physics,
Astro-Particle Theory \& Cosmology Group, Hounsfield Road, Hicks
Building, University of Sheffield, Sheffield S3 7RH, UK}
\address{$^3$ Department of Applied Mathematics,
Astro-Particle Theory \& Cosmology Group, Hounsfield Road, Hicks
Building, University of Sheffield, Sheffield S3 7RH, UK}
\address{$^4$ Institute for Theoretical Physics, University of
Heidelberg, D-69120 Heidelberg, Germany}
\address{$^5$ Institute of Theoretical Astrophysics, University of
Astrophysics, N-0315 Oslo, Norway}
\address{$^6$ Deutsches Elektron-Synchroton DESY, Hamburg, Notkestr. 85, 22607 Hamburg, Germany}
\address{$^7$ Department of Physics and Astronomy, The Johns Hopkins University,
Baltimore, MD 21218, USA}

\eads{\mailto{oeb@phys.au.dk}, \mailto{php04awb@sheffield.ac.uk},
\mailto{C.vandebruck@sheffield.ac.uk}, \mailto{sth@phys.au.dk},
\mailto{d.mota@thphys.uni-heidelberg.de}, \mailto{lily.schrempp@desy.de},
\mailto{dtv@skysrv.pha.jhu.edu}}
\date{{\today}}

\begin{abstract}
A coupling between a light scalar field and neutrinos has been
widely discussed as a mechanism for linking (time varying) neutrino
masses and the present energy density and equation of state of dark
energy. However, it has been pointed out that the viability of this
scenario in the non-relativistic neutrino regime is threatened by
the strong growth of hydrodynamic perturbations associated with a
negative adiabatic sound speed squared. In this paper we revisit the
stability issue in the framework of linear perturbation theory in a
model independent way. The criterion for the stability of a model is
translated into a constraint on the scalar-neutrino coupling, which
depends on the ratio of the energy densities in neutrinos and cold
dark matter. We illustrate our results by providing meaningful
examples both for stable and unstable models.

\end{abstract}

\maketitle

\section{Introduction} 

Precision observations of the cosmic microwave background
~\cite{Bennett:2003bz,Spergel:2003cb,Spergel:2006hy}, the large scale
structure of galaxies \cite{Tegmark:2006az}, and distant type Ia supernovae
\cite{Riess:1998cb,Perlmutter:1998np,Astier:2005qq,Wood-Vasey:2007jb} have led
to a new standard model of cosmology in which the energy density is dominated
by dark energy with negative pressure, leading to an accelerated expansion of
the universe.

The simplest possible explanation for dark energy is the cosmological constant
which has $P=w\rho$ with $w=-1$ at all times. However, since the cosmological
constant has a magnitude completely different from theoretical expectations
one is naturally led to consider other explanations for the dark energy.  A
light scalar field rolling in a very flat potential would for instance be a
candidate better motivated from high energy physics
~\cite{Wetterich:1987fm,Peebles:1987ek,Ratra:1987rm}. In the limit of a
completely flat potential it would have $w=-1$. Such models are generically
known as quintessence models
\cite{Zlatev:1998tr,Wang:1999fa,Steinhardt:1999nw,Barreiro:1999zs,Baccigalupi:2001aa,%
  Caldwell:2003vp}. The scalar field is usually assumed to be minimally
coupled to matter and to curvature, but very interesting effects can occur if
this assumption is relaxed (see for
instance~\cite{Mota:2004pa,Amendola:1999er,domenico:2002,domenico,Bertolami:1999dp,tomi,Perrotta:1999am}).
In general such models alleviate the required fine tuning in order to achieve
$\Omega_X \sim \Omega_m$, where $\Omega_X$ and $\Omega_m$ are the dark energy
and matter densities at present. Also by properly choosing the quintessence
potential it is possible to achieve tracking behaviour of the scalar field so
that one also avoids the extreme fine tuning of the initial conditions for the
field.

Many other possibilities have been considered, like $k$-essence,
which is essentially a scalar field with a non-standard kinetic term
~\cite{Armendariz-Picon:1999rj,Chiba:1999ka,Armendariz-Picon:2000ah}.
It is also possible, although not without problems, to construct
models which have $w<-1$, the so-called phantom dark energy models
~\cite{Caldwell:1999ew,Schulz:2001yx,Carroll:2003st}. Finally, there
are even more exotic models where the cosmological acceleration is
not provided by dark energy, but rather by a modification of the
Friedmann equation due to modifications of gravity on large
scales~\cite{Deffayet:2001pu,Dvali:2003rk}, or even due to higher
order curvature terms in the gravity Lagrangian
\cite{sean,brook,morad}.

A very interesting proposal is the so-called mass varying neutrino
(MaVaN) model~\cite{Hung:2000yg,Gu:2003er,Fardon:2003eh} in which a
light scalar field couples to neutrinos. Due to the coupling, the mass
of the scalar field does not have to be as small as the Hubble scale
but can be much larger, while the model still accomplishes late-time
acceleration. This scenario also holds the interesting possibility of
circumventing the well-known cosmological bound on the neutrino mass
\cite{Hannestad:2006zg,Lesgourgues:2006nd,Hannestad:2003xv,Hannestad:2005gj,Goobar:2006xz,Spergel:2006hy,%
Seljak:2006bg,Feng:2006zj,Hannestad:2006mi,Fogli:2006yq,Tegmark:2006az,skordis,Zunckel:2006mt,Elgaroy:2006iy}.
The scenario is a variant of the chameleon cosmology
model~\cite{Khoury:2003aq,Brax:2004qh,shaw2} in which a light scalar
field couples democratically to all non-relativistic matter.

The idea in the MaVaN model is to write down an effective potential
for the scalar field which as a result of the coupling contains a term
related to the neutrino energy density. If the pure scalar field
potential is tuned appropriately the effective potential including the
neutrino contribution will have a minimum with a steep second
derivative for some finite scalar field VEV. The scalar field is
therefore locked in the minimum and when the minimum evolves due to
changing neutrino energy density the field tracks this evolution
adiabatically. This naturally leads to a dynamical effective equation
of state for the combined scalar - neutrino fluid close to $w=-1$
today, and to a neutrino mass which is related to
the combined neutrino-scalar field fluid's energy density $\rho_{\rm
DE}$. Since $\rho_{\rm DE}$ decreases with time, also the neutrino
mass varies in this kind of scenario, where its present value is
explained in terms of $\rho_{\rm DE}^{1/4}(a=1)$. Possible tests for
the MaVaN scenario can be found in
Ref.~\cite{Kaplan:2004dq,Barger:2005mn,Cirelli:2005sg,Barger:2005mh,%
Gu:2005pq,Li:2004tq,Ringwald:2006ks,Schrempp:2006mk}.

MaVaN models, however, suffer from the problem that for some choices
of scalar-neutrino couplings and scalar field potentials the combined
fluid is subject to an instability once the neutrinos become
non--relativistic. Effectively the scalar field mediates an attractive
force between neutrinos which can possibly lead to the formation of
neutrino nuggets~\cite{Afshordi:2005ym}. This in turn would make the
combined fluid behave like cold dark matter and thus render it
non-viable as a candidate for dark energy.

In perturbation theory the formation of these nuggets can be seen as
a consequence of an imaginary speed of sound for the combined fluid,
signaling fast growth of instabilities. However, an imaginary speed
of sound cannot be generally used as a sufficient criterion for the
instabilities, as the drag provided by cold dark matter may postpone
those instabilities.

The instability can possibly occur in these models because the
effective mass associated with the scalar field is much larger than
$H$. Accordingly, on sub-Horizon scales larger than the effective
Compton wavelength of the scalar field $m_{\phi}^{-1}<a/k<H^{-1}$
the perturbations are adiabatic.

This is a consequence of the steepness of the effective potential
and can be remedied by making the potential sufficiently flat. In
this case the evolution of the field is highly non-adiabatic
\cite{Brookfield:2005td,Brookfield:2005bz}.  However, this model has
the disadvantage that the neutrino mass is no longer related
naturally to the dark energy density and equation of state.

In this paper we study various choices of scalar-neutrino couplings
and scalar field potentials with the aim of identifying the
conditions for the instability to occur. In the next section we
review the formalism needed to study mass varying neutrinos and in
section 3 we derive the equation of motion of the neutrino
perturbations. Section 4 contains our results for various couplings
and potentials, and finally section 5 contains a discussion and
conclusion.

\section{\label{sec:formalism}Formalism} 

The idea in the so-called Mass Varying Neutrino (MaVaN)
scenario~\cite{Hung:2000yg,Gu:2003er,Fardon:2003eh} is to introduce
a coupling between (relic) neutrinos and a light scalar field and to
identify this coupled fluid with dark energy. As a direct
consequence of this new interaction, the neutrino mass $m_\nu$ is
generated from the vacuum expectation value (VEV) of the scalar
field and becomes linked to its dynamics. Thus the pressure
$P_\nu(m_\nu(\phi),a)$ and energy density $\rho_\nu(m_\nu(\phi),a)$
of the uniform neutrino background contribute to the effective
potential $V(\phi,a)$ of the scalar field.  The effective potential is defined by
\begin{equation}
 \centering
 V(\phi) = V_\phi(\phi)+(\rho_\nu-3 P_\nu)
 \label{eq:eff}
\end{equation}
where $V_\phi(\phi)$ denotes the fundamental scalar potential and
$a$ is the scale factor. Throughout the paper we assume a flat
Friedman-Robertson-Walker cosmology and use the convention $a_0=1$,
where we take the subscript $0$ to denote present day values.

Assuming the neutrino distribution to be Fermi-Dirac and
neglecting the chemical potential, the energy density and pressure
of the neutrinos can be expressed in the following form
\cite{Peccei:2004sz}
\begin{eqnarray}
 \centering
\label{eq:inte}
 \rho_\nu(a,\phi)&=&\frac{T_\nu^4(a)}{\pi^2}\int_0^\infty
 \frac{dy\,y^2\sqrt{y^2+\frac{m_\nu^2(\phi)}{T_\nu^2(a)}}}{e^{y}+1},\nonumber
 \\
P_\nu(a,\phi)&=&\frac{T_\nu^4(a)}{3\pi^2}\int_0^\infty
 \frac{dy\,y^4}{\sqrt{y^2+\frac{m_\nu^2(\phi)}{T_\nu^2(a)}}(e^{y}+1)},
\end{eqnarray}
where $T_\nu=T_{\nu_0}/a$ is the neutrino temperature and $y$ corresponds
to the ratio of the neutrino momentum and neutrino temperature,
$y=p_\nu/T_\nu$.

The energy density and pressure of the scalar field are given by the usual expressions,
\begin{eqnarray}
 \centering
 \rho_\phi(a)&=&\frac{1}{2a^2}\dot{\phi}^2+V_\phi(\phi),\nonumber\\
P_\phi(a)&=&\frac{1}{2a^2}\dot{\phi}^2-V_\phi(\phi).
\end{eqnarray}

Defining $w=P_{\rm DE}/\rho_{\rm DE}$ to be the equation of state of
the coupled dark energy fluid, where $P_{\rm DE}=P_\nu+P_\phi$
denotes its pressure and $\rho_{\rm DE}=\rho_\nu+\rho_\phi$ its
energy density, and the requirement of energy conservation gives,
\begin{equation}
\centering \dot{\rho}_{DE}+3H\rho_{DE}(1+w)=0. \label{eq:Econserv}
\end{equation}
Here $H\equiv \frac{\dot{a}}{a}$ and we use dots to refer to the
derivative with respect to conformal time. Taking
Eq.~(\ref{eq:Econserv}) into account, one arrives at a modified
Klein-Gordon equation describing the evolution of $\phi$,
\begin{equation}
\ddot{\phi}+2H\dot{\phi}+a^2 V_\phi^{\prime}=-a^2\beta(\rho_\nu-3p_\nu).
\label{eq:KG}
\end{equation}
Here and in the following
primes denote derivatives with respect to $\phi$ ($^\prime =
\partial/\partial \phi$) and $\beta=\frac{d {\rm log}m_\nu}{d\phi}$
is the coupling between the scalar field and the neutrinos.

\subsection{The fully adiabatic case}

In the following let us consider the late time evolution of the
coupled scalar-neutrino fluid in the limit $m_\nu\gg T_\nu$ where
the neutrinos are non-relativistic. It is in this regime that MaVaN
models can potentially become unstable for the following reason: The
attractive force mediated by the scalar field (which can be much
stronger than gravity) acts as a driving force for the
instabilities. But as long as the neutrinos are still relativistic,
the evolution of the density perturbations will be dominated by
pressure which inhibits their growth, as the strength of the
coupling is suppressed when $\rho_\nu = 3P_\nu$.

In the non-relativistic limit $m_\nu\gg T_\nu$, the expressions for
the energy density and pressure in neutrinos in Eq.~(\ref{eq:inte})
reduce to
\begin{eqnarray}
 \centering \rho_\nu&\simeq& m_\nu n_\nu,\nonumber\\ P_\nu&\simeq&
 0,\label{eq:limit}
\end{eqnarray}
such that Eq.~(\ref{eq:eff}) takes the form
\begin{equation}
V=\rho_\nu+V_\phi=m_\nu n_\nu+V_\phi.\label{eq:effNR}
\end{equation}

Assuming the curvature scale of the potential and thus the mass of the
scalar field $m_\phi$ to be much larger than the expansion rate of the
Universe,
\begin{equation}
\label{eq:mphi}
V^{\prime\prime}=\rho_\nu\left(\beta^{\prime}+\beta^2\right)+V_\phi^{\prime\prime}\equiv
m_\phi^2\gg H^2,
\end{equation}
the adiabatic solution to the equation of motion of the scalar field
in Eq.~(\ref{eq:KG}) applies \cite{Fardon:2003eh}\footnote{In this
case for $|\phi|< M_{\rm pl}\simeq 3\times 10^{18}$ GeV the effects of
the kinetic energy terms can be safely
ignored~\cite{Fardon:2003eh}.}. As a consequence, the scalar field
instantaneously tracks the minimum of its effective potential $V$,
solution to the condition
\begin{equation}
 \centering
 V^{\prime}=\rho_\nu^{\prime}+V_\phi^{\prime}=m_\nu^{\prime}\left(\frac{\partial
 \rho_\nu}{\partial m_\nu}+\frac{\partial V_\phi}{\partial
 m_\nu}\right)=m_\nu^{\prime}\left(n_\nu+\frac{\partial
 V_\phi}{\partial m_\nu}\right)=0.
 \label{eq:phi}
\end{equation}
As the universe expands the neutrino energy density gets diluted, thus
naturally giving rise to a slow evolution of $V(\phi)$. Consequently,
the value of the scalar field $\phi$ evolves on cosmological time
scales. Note that as long as $m_\nu^{\prime}$ does not vanish, this
implies that also the neutrino mass $m_\nu(\phi)$ is promoted to a
time dependent, dynamical quantity. Its late time evolution can be
determined from the last equality in Eq.~(\ref{eq:phi}).

In order to specify good candidate potentials $V_{\phi}(\phi)$ for a
viable MaVaN model of dark energy, we must demand that the equation
of state parameter $w$ of the coupled scalar-neutrino fluid today
roughly satisfies $w\sim-1$ as suggested by observations
\cite{Tonry:2003zg}. By noting that for constant $w$ at late times,
\begin{equation}
\rho_{\rm DE}\sim V \propto a^{-3(1+w)}
\end{equation}
and by requiring energy conservation according to
Eq.~(\ref{eq:Econserv}), one arrives at \cite{Fardon:2003eh}
\begin{equation}
 \centering
 1+w=-\frac{1}{3}\frac{\partial \log\, V }{\partial \log
 a}.
\end{equation}
In the non-relativistic limit $m_\nu\gg T_\nu$ this is equivalent to
\begin{equation} \label{eq:eos}
 \centering
1+w=-\frac{a}{3V }\left(m_\nu\frac{\partial n_\nu}{\partial
a}+n_\nu\frac{\partial m_\nu}
 {\partial a}+\frac{V^{\prime}_\phi}{a^{\prime}}\right)=-\frac{m_\nu V^{\prime}_\phi}{ m^{\prime}_\nu V},
\end{equation}
where in the last equality it has been used that $V^{\prime}=0$
according to Eq.~(\ref{eq:phi}). To allow for an equation of
state close to $w\sim -1$ today one can conclude that either the
scalar potential $V_\phi$ has to be fairly flat or the dependence of
the neutrino mass on the scalar field has to be very steep.

\subsection{\label{sec:general}The general case}

As it will turn out later, the influence of the cosmic expansion in
combination with the gravitational drag exerted by CDM on the
neutrinos can have an effect on the stability of a MaVaN model.
However, to begin we will neglect any growth-slowing effects on the
perturbations and proceed with a more general analysis of this case.
Under these circumstances, the dynamics of the perturbations are solely
determined by the sound speed squared which for a general fluid
component $i$ takes the following form,
\begin{equation}
c_{s i}^2=\frac{\delta P_i}{\delta \rho_i},
\label{eq:generalsound}
\end{equation}
where $P_i$ and $\rho_i$ denote the fluid's pressure and energy
density, respectively. The sound speed $c_{s i}^2$ can be
expressed in terms of the sound speed $c_{a i}^2$ arising from
purely adiabatic perturbations as well as an additional entropy
perturbation $\Gamma_i$ and the density contrast $\delta_i=\delta
\rho_i/\rho_i$ in the given frame~\cite{Bean:2003fb,Hannestad:2005ak},
\begin{eqnarray}
w_i\Gamma_i&=&(c_{s i}^2 -c_{a i}^2)\,\delta_i,\hspace{2ex}\\
&=&\frac{\dot{P}_i}{\rho_i}\left(\frac{\delta
P_i}{\dot{P}_i}-\frac{\delta \rho_i}{\dot{\rho}_i}\right).
\label{eq:Gamma}
\end{eqnarray}
Here $w_i$ denotes the equation of state parameter and $\Gamma_i$ is
a measure for the relative displacement between hypersurfaces of
uniform pressure and uniform energy density. For most dark energy
candidates (like quintessence or k-essence) dissipative processes
evoke entropy perturbations and thus $\Gamma_i\neq 0$.

However, in MaVaN models the effective mass of the scalar field
$m_{\phi} \gg H$ sets the scale, $m_{\phi}^{-1}$, where these
processes and the associated gradient terms become
unimportant~\cite{Afshordi:2005ym,Kaplinghat:2006jk}, to be much
smaller than the Hubble radius (in contrast to a quintessence field
with finely-tuned mass $\lesssim H$ and long range $\gtrsim
H^{-1}$). As a consequence, on sub-Hubble scales
$H^{-1}>\frac{a}{k}>m_\phi^{-1}$ all dynamical properties of
(non-relativistic) MaVaNs are set by the local neutrino energy
density~\cite{Afshordi:2005ym}. In particular, for small deviations
away from the minimum of its effective potential, the scalar field
re-adjusts to its new minimum on time scales $\sim m_\phi^{-1}$
small compared to the characteristic cosmological time scale
$H^{-1}$. In this case the hydrodynamic perturbations in MaVaNs are
adiabatic. This means the system of neutrinos and the scalar field
can be treated as a unified fluid with pressure
$P_{DE}=P_\nu+P_\phi$ and energy density
$\rho_{DE}=\rho_\nu+\rho_\phi$ without intrinsic entropy,
$\Gamma_{DE}=0$ \footnote{(see also \cite{pedro} for another example of unified models)}.

If any growth-slowing effects can be neglected, the perturbations in
a MaVaN model are driven by the effective sound speed squared given
by
\begin{equation}
c_a^2=\frac{\dot{P}_{\rm
DE}}{\dot{\rho}_{DE}}=\frac{\dot{w}\rho_{DE}+w\dot{\rho}_{DE}}{\dot{\rho}_{DE}}=w-\frac{\dot{w}}{3H(1+w)}
, \label{eq:adiabatic}
\end{equation}
where Eq.~(\ref{eq:Econserv}) and Eq.~(\ref{eq:Gamma}) have been
used. In the case $c_a^2>0$ the attractive scalar force is offset by
pressure forces and the fluctuations oscillate as sound waves and
can be considered as stable. However, for $c_a^2<0$ perturbations
become unstable and tend to blow up.

\section{\label{sec:Perturbations}Evolution of the Perturbations}

In this section we will analyse the linear MaVaN perturbations in
the synchronous gauge, which is characterised by a perturbed line
element of the form
\begin{equation}
ds^2=a(\tau)^2(-d\tau^2+(\delta_{ij}+h_{ij})dx^{i}dx^j),
\end{equation}
where $\tau$ denotes conformal time and $h_{ij}$ is the metric
perturbation. Here and in the following dots represent derivatives
with respect to $\tau$. Most of our other notations and conventions
comply with those in Ma and Bertschinger~\cite{mb}. Consequently,
the Friedmann equation takes the form
\begin{equation}
3H^2=\frac{a^2}{M_{\rm pl}^2}\left(\frac{\dot{\phi}^2}{2a^2}+V_\phi(\phi)+\rho_m\right),
\end{equation}
with $M_{\rm pl}\equiv(\sqrt{8\pi G})^{-1}$ denoting the reduced
Planck mass and the subscript $m$ comprising all matter species.

Since the following perturbation equations have been widely
discussed in the literature
(e.g.~\cite{Amendola:2003wa,domenico:2002,Brax:2004qh,Koivisto:2005nr,Brookfield:2005bz}
and references therein), we will simply state them here for
neutrinos coupled to a scalar field.

The evolution equation for the MaVaN density contrast
$\delta_\nu=\delta\rho_\nu/\rho_\nu$ is given
by~\cite{Brookfield:2005bz},
\begin{eqnarray}\label{denscont}
\dot{\delta}_\nu &=& 3\left(H+\beta
\dot{\phi}\right)\left(w_\nu-\frac{\delta
p_\nu}{\delta\rho_\nu}\right)\delta_\nu -
\left(1+w_\nu\right)\left(\theta_\nu +
\frac{\dot{h}}{2}\right)\nonumber \\ &+&
\beta\left(1-3w_\nu\right)\delta\dot{
  \phi}+\beta^{\prime}\dot{\phi}\delta\phi\left(1-3w_\nu\right),
\end{eqnarray}
where $\beta=\frac{d\log m_\nu}{d \phi}$.

Furthermore, the trace of the metric perturbation, $h\equiv
\delta^{ij}h_{ij}$, according to the linearised Einstein equations
satisfies,
\begin{eqnarray}\label{eomh}
\ddot{h} +H\dot{h}&=&\frac{a^2}{M_{\rm pl}^2}[\delta T^{0}_0-\delta
T^{i}_i],\,\,\mbox{where}\\ \delta
T^{0}_0&=&-\frac{1}{a^2}\dot{\phi}\delta\dot{\phi}-V_{\phi}^{\prime}(\phi)\delta\phi-\sum\limits_{m}^{}\rho_m\delta_m,\\
\delta
T^{i}_i&=&\frac{3}{a^2}\dot{\phi}\delta\dot{\phi}-3V_{\phi}^{\prime}(\phi)\delta
\phi+\sum
\limits_{r}^{}\rho_{r}\delta_{r}+3c_b^2\rho_b\delta_b+3c_\nu^2\rho_\nu\delta_\nu.
\end{eqnarray}
Here $\delta T^{\mu}_\nu$ denotes the perturbed stress energy tensor
and the subscripts $m$ and $r$ collect neutrinos, radiation, CDM and
baryons (with sound speed $c_b$) as well as (relativistic) neutrinos
and radiation, respectively.

The evolution equation for the neutrino velocity perturbation
$\theta_\nu\equiv i k_i v^{i}_\nu$ with $v^{i}_\nu\equiv
dx^{i}/d\tau$ reads~\cite{Brookfield:2005bz},
\begin{eqnarray}\label{velpert}
\dot{\theta}_\nu &=& -H(1-3w_\nu)\theta_\nu -
\frac{\dot{w}_\nu}{1+w_\nu}\theta_\nu+\frac{\frac{\delta
p_\nu}{\delta\rho_\nu}}{1+w_\nu}k^{2}\delta_\nu \nonumber \\ &+&
\beta\frac{1-3w_\nu}{1+w_\nu} k^{2} \delta
\phi-\beta(1-3w_\nu)\,\dot{\phi}\,\theta_\nu - k^{2} \sigma_\nu,
\end{eqnarray}
where $\sigma_\nu$ denotes the neutrino shear as defined in
~\cite{mb}.

Finally, the perturbed Klein-Gordon equation for the coupled scalar
field is given by~\cite{Brookfield:2005bz}
\begin{eqnarray}\label{pertkg}
\lefteqn{\ddot{\delta \phi}+ 2H \dot{\delta \phi}+\left[k^{2}
+a^2\left\{V_{\phi}^{\prime\prime}+\beta^{\prime}(\rho_\nu-3
P_\nu)\right\}\right]\delta \phi+\frac{1}{2}\dot{h} \dot{\phi}=}
\\ \nonumber & &-a^2\beta\delta_\nu \rho_\nu(1-3 \frac{\delta
p_\nu}{\delta\rho_\nu}).
\end{eqnarray}

We note that instead of proceeding via the fluid equations,
Eqs.~(\ref{denscont}) and (\ref{velpert}), the evolution of the
neutrino density contrast can be calculated from the Boltzmann
equation \cite{mb}. We have verified analytically and numerically
that the two methods yield identical results provided that the
scalar-neutrino coupling is appropriately taken account of in the
Boltzmann hierarchy~\cite{Keum}.

As discussed in sec.~\ref{sec:formalism} MaVaNs models can only possibly
become unstable on sub-Hubble scales $m_\phi^{-1}<a/k<H^{-1}$ in the
non-relativistic regime of the neutrinos, where the perturbations evolve
adiabatically. For our numerical results in the next section we solve the
coupled Eqs.~(\ref{denscont}-\ref{pertkg}) in the (quasi-)adiabatic regime by
neglecting the neutrino shear $\sigma_\nu$. This approximation is justified,
since the scalar-neutrino coupling becomes important in this regime and
$m_\nu$ is much larger than the mean momentum of the neutrino distribution.

For the purpose of gaining further analytical insight into the evolution of
the neutrino density contrast, it is instructive to apply additional
approximations to Eqs.~(\ref{denscont}-\ref{pertkg}) to be justified in the
following.

Since the minimum of the effective potential tracked by the scalar
field evolves only slowly due to changes in the neutrino energy
density, we can safely ignore terms proportional to $\dot{\phi}$.
Moreover, in the non-relativistic regime of the neutrinos on scales
$m_\phi^{-1}<a/k<H^{-1}$, as a consequence of $P_\nu\sim 0$ it
follows that $\sigma_\nu\sim0$ and $w_\nu \sim
0$ as well as $\rho_r\sim c_b^2\sim 0$. In addition, in the
following we substitute $\delta \phi$ by its average value
corresponding to the forcing term on the right hand side of
Eq.~(\ref{pertkg}) in the above limits,
\begin{equation}\label{averagedphi}
\delta\bar{\phi}=-\frac{\beta\rho_\nu\delta_\nu}{(V_{\phi}^{\prime\prime}+\rho_\nu\beta^{\prime})+\frac{k^2}{a^2}},
\end{equation}
which solves the perturbed Klein-Gordon equation reasonably well on
all scales~\cite{domenico:2002,Koivisto:2005nr}. Finally, by combining
the derivative of Eq.~(\ref{denscont}) with Eq.~(\ref{eomh}) --
Eq.~(\ref{velpert}) and Eq.~(\ref{averagedphi}) in the
non-relativistic limit, we arrive at the equation of motion for the
neutrino density contrast valid at late times on length scales
$m_\phi^{-1}<a/k<H^{-1}$,
\begin{eqnarray}\label{denseom}
\hspace{-2.5cm}\ddot{\delta}_\nu
+H\dot{\delta}_\nu+\left(\frac{\delta
p_\nu}{\delta\rho_\nu}k^2-\frac{3}{2}H^2\Omega_{\nu}\frac{G_{\rm
eff}}{G}\right)\delta_\nu=
\frac{3}{2}H^2\left[\phantom{\frac{|}{|}}\Omega_{\rm CDM}\delta_{\rm
CDM}+\Omega_{b}\delta_{b}\phantom{\frac{|}{|}}\right]\nonumber\\\nonumber \\
\end{eqnarray}
where
\begin{eqnarray}
G_{\rm eff}&=&G\left(1+\frac{2\beta^2 M^2_{\rm
pl}}{1+\frac{a^2}{k^2}
\{V_{\phi}^{\prime\prime}+\rho_\nu\beta^{\prime}\}}\right)\,\mbox{and}\label{Geff}\\
\Omega_i&=&\frac{a^2\rho_i}{3H^2M^2_{\rm pl}}.
\end{eqnarray}
Since neutrinos not only interact through gravity, but also through
the force mediated by the scalar field, they feel an effective
Newton's constant $G_{\rm eff}$ as defined in Eq.~(\ref{Geff}). The
force depends upon the MaVaN model specific functions $\beta$ and
$V_{\phi}$ and takes values between $G$ and $G(1+2\beta^2 M^2_{\rm
pl})$ on very large and small length scales, respectively. The scale
dependence of $G_{\rm eff}$ is due to the finite range of the scalar
field
$(V_{\phi}^{\prime\prime}+\rho_\nu\beta^{\prime})^{-\frac{1}{2}}$,
which according to Eq.~(\ref{eq:mphi}) is equal to
$(m_{\phi}^2-\beta^2\rho_\nu)^{-\frac{1}{2}}$. For moderate coupling
strength it is essentially given by the inverse scalar field mass,
whereas for $\beta \gg 1/M_{\rm pl}$ it can take larger values.
Accordingly, in a MaVaN model both the scalar potential $V_\phi$ and
the coupling $\beta$ influence the range of the scalar field force
felt by neutrinos, whereas its strength is determined by the
coupling $\beta$.

The evolution of perturbations in cold dark matter (CDM) coupled to
a light scalar field in coupled quintessence ~\cite{domenico:2002}
and chameleon cosmologies~\cite{Brax:2004qh} is governed by an
equation similar to Eq.~(\ref{denseom}). However, we would like to
point out that for the same coupling functions the dynamics of the
perturbations in neutrinos can be quite different from those in
coupled CDM. This is a result of the fact that $\Omega_{\nu}\ll
(\Omega_{\rm CDM}+\Omega_{\rm b})$. Whereas $\Omega_{\rm CDM}\sim
0.2$ and $\Omega_{\rm b}\sim 0.05$~\cite{Tegmark:2006az} at present,
$\Omega_{\nu}$ depends on the so far not known absolute neutrino
mass scale realised in nature. Taking as a lower bound the mass
splitting deduced from atmospheric neutrino flavour oscillation
experiments and the upper bound derived from the Mainz tritium
beta-decay experiments~\cite{kraus}, we get $10^{-4}\lesssim
\Omega_{\nu}\lesssim 0.15$ today \footnote{Note that if the upper
limit from the Mainz experiment is saturated the requirement
$\Omega_\nu \ll \Omega_m$ is formally not satisfied. However, this
case should be viewed as very extreme and is most likely excluded
based on structure formation arguments}. It is important to note
that since in the standard MaVaN scenario the neutrino mass is an
increasing function of time, at earlier times the ratio
$\Omega_{\nu}/(\Omega_{\rm CDM}+\Omega_{\rm b})$ was even smaller
than today. In general it follows that the smaller this ratio is,
the larger the relative influence of the forcing term on the RHS of
Eq.~(\ref{denseom}) becomes. The forcing term describes the effect
of the perturbations of other cosmic components on the dynamics of
the neutrino density contrast and competes with the scalar field
dependent term $\propto \frac{G_{\rm eff}}{G}\Omega_{\nu}\delta_\nu$
on the LHS. Correspondingly, apart from the scalar field mediated
force the neutrinos feel the gravitational drag exerted by the
potential wells formed by CDM. Consequently, as long as the coupling
function $\beta$ does not considerably enhance the influence of the
term $\propto \frac{G_{\rm eff}}{G}\Omega_{\nu}\delta_\nu$, the
non-relativistic neutrinos will follow CDM (like baryons) just as in
the Standard Model.

In the following we classify the behaviour of the neutrino density
contrast in models of neutrino dark energy subject to all relevant
kinds of coupling functions $\beta$. We emphasize that this
classification is completely model independent. In the small-scale
limit we distinguish the following three cases:

\textbf{Small-scale limit}
\begin{itemize}
\item[a)] For $\Omega_{\nu}(1+2\beta^2M_{\rm pl}^2) < \Omega_{\rm CDM}$
until the present time, the neutrino density contrast is stabilised
by the CDM source term which dominates its dynamics. In this case
the influence of the scalar field on the perturbations is
subdominant and the density contrast in MaVaNs grows moderately just
like gravitational instabilities in uncoupled neutrinos.

\item[b)] For $\beta\sim {\rm const.}$ and much larger than all other parameters at
late times, $\Geff \gg G$, the damping term
$H\dot{\delta}_\nu$ in Eq.~(\ref{denseom}) as well as the the terms
proportional to $\delta_{\rm CDM}$ and $\delta_b$ can be neglected,
leading to exponentially growing solutions.
\item[c)] For $\beta\neq {\rm const.}$ and growing faster than all other parameters at
late times, $\Geff \gg G$, $\delta_\nu$ is growing faster than
exponentially\footnote{In the limit $\beta(\tau)\rightarrow\infty$
for $\tau\rightarrow\infty$, Eq.~\ref{denseom} takes the form
$\ddot{\delta}_\nu-3H^2\Omega_\nu\frac{\beta^2(\tau) M^2_{\rm
pl}}{1+a^2(V_{\phi}^{\prime\prime}+\rho_\nu\beta^{\prime})/k^2}\delta_\nu=0$,
and it can be shown that
$|\frac{\dot{\delta}_\nu}{\delta_\nu}|\rightarrow \infty$ for
$\tau\rightarrow\infty$~\cite{Kamke}. Since this ratio is constant
and thus not large enough for an exponentially growing $\delta_\nu$,
the solution is required to grow faster than exponentially.}.
\end{itemize}
\smallskip

\noindent{In} contrast, on scales
$({V_{\phi}^{\prime\prime}+\rho_\nu\beta^{\prime}})^{-1/2}\ll
a/k<H^{-1}$ much larger than the range of the $\phi$-mediated force,

\textbf{Large-scale limit}
\begin{itemize}
\item[d)] For $\beta\sim {\rm const.}$ and of moderate strength, $G_{\rm
eff}\sim G$ and the perturbations behave effectively like
perturbations for uncoupled fluids in General Relativity.
\item[e)] For $\beta$ growing faster than all other quantities at
late times, $G_{\rm eff} \gg G$, instabilities develop on all
sub-Hubble scales
$a/k>({V_{\phi}^{\prime\prime}+\rho_\nu\beta^{\prime}})^{-1/2}$
according to $c)$. However, on large length scales their growth rate
is suppressed due to the corresponding small wave number $k$.
\end{itemize}

\subsection{\label{sec:Potentials}Potentials and Couplings}

In the following, we consider two combinations of scalar potentials
$V_\phi(\phi)$ and of scalar-neutrino couplings $\beta$ which define
our MaVaN models. The potentials are chosen to accomplish the
required cosmic late-time acceleration and for the couplings we take
meaningful limiting cases.

Our main point is to present a proof of concept of the stability
conditions stated above, which is valid for a general adiabatic
MaVaN model. We note that a certain degree of fine-tuning is
exerted. It is mainly due to the fact that CMBFAST and CAMB only
operate in the linear regime $H\sim10^{-4}{\rm Mpc}^{-1}<k<0.1{\rm
Mpc}^{-1}$ and correspondingly, only in this regime can we
analytically track the evolution of perturbation by the help of
linear theory. Since the Compton wavelength of the scalar field
$\sim m^{-1}_{\phi}$ sets the length scale of interest where
possible instabilities can grow fastest, $a/k\gwig m_{\phi}^{-1}$,
(cf. the discussion in sec.~\ref{sec:formalism}), this implies the
scalar field mass has to be tuned accordingly, while at the same
time the correct cosmology has to be accomplished.

\begin{figure}
\begin{center}
\includegraphics*[width=10cm]{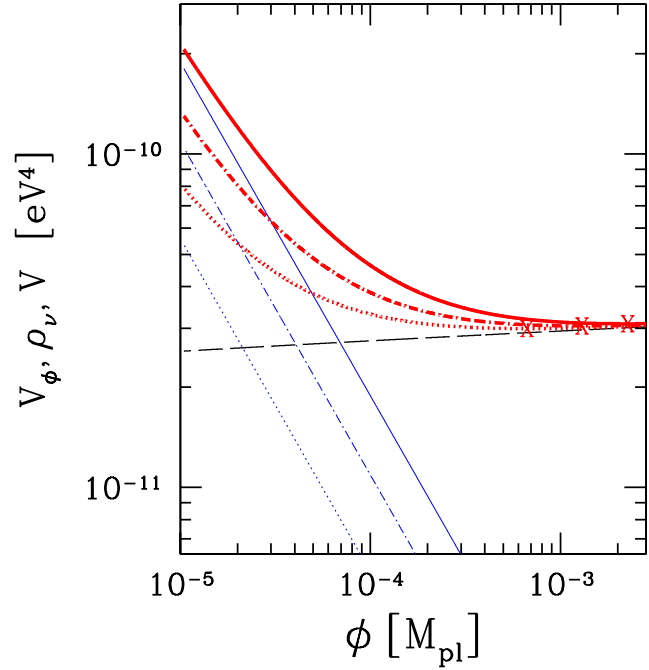}
\caption[]{The effective potential $V$ (thick lines), composed of
the scalar potential $V_\phi$ (dashed) and the neutrino energy
density $\rho_\nu$, plotted for three different redshifts, $z=5$
(solid), $z=4$ (dashed-dotted), $z=3$ (dotted). The VEV of $\phi$
tracks the minimum of $V$ (marked by X) and evolves to smaller
values for decreasing redshift. We have used
$\kappa=1\times10^{20} M_{\rm pl}^{-1}$ and
$V_0=8.1\times10^{-13}$eV$^4$.
}
\label{fig:fig3}
\end{center}
\end{figure}

Firstly, we consider a MaVaN model suggested by~\cite{Fardon:2003eh}
which we will refer to as the \emph{log-linear model}. The scalar
field has a Coleman-Weinberg type~\cite{Coleman:1973jx} logarithmic
potential,
\begin{equation}
 \centering V_\phi(\phi)=V_0\,\log(1+\kappa\phi),
 \label{eq:lln}
\end{equation}
where the constants $V_0$ and $\kappa$ are chosen appropriately to
yield $\Omega_{\rm DE}\sim 0.7$ and $m_\phi\gg H$ today. The choice
of $V_{\phi}$ determines the evolution of $\phi$ according to
Eq.~(\ref{eq:effNR}) as plotted in fig.~\ref{fig:fig3}. Apparently,
the neutrino background has a stabilising effect on $\phi$. It
drives the scalar field to larger values and stops it from rolling
down its potential $V_\phi$. This competition of the two terms in
Eq.~(\ref{eq:effNR}) results in a minimum at an intermediate value
of $\phi$ (cf. Eq.~\ref{eq:phi}), which slowly evolves due to
changes in the neutrino energy density. As the universe expands and
$\rho_\nu$ dilutes, both the minimum and the scalar field are driven
to smaller values towards zero.

Let us now turn to the neutrino mass and its evolution. The
dependence of $m_\nu$ on the scalar field is given by,
\begin{equation}
 \centering
 m_\nu(\phi)=m_0\frac{\phi_0}{\phi}.
 \label{eq:masslln}
\end{equation}
Such a dependence naturally emerges in the framework of the seesaw
mechanism. In this case the light neutrino mass $m_\nu$ arises from
integrating out a heavier sterile state, whose mass varies linearly
with the value of the scalar field (as e.g. in
Ref.~\cite{Fardon:2003eh,Afshordi:2005ym,Spitzer:2006hm}).

\begin{figure}
\begin{center}
\includegraphics*[width=8cm]{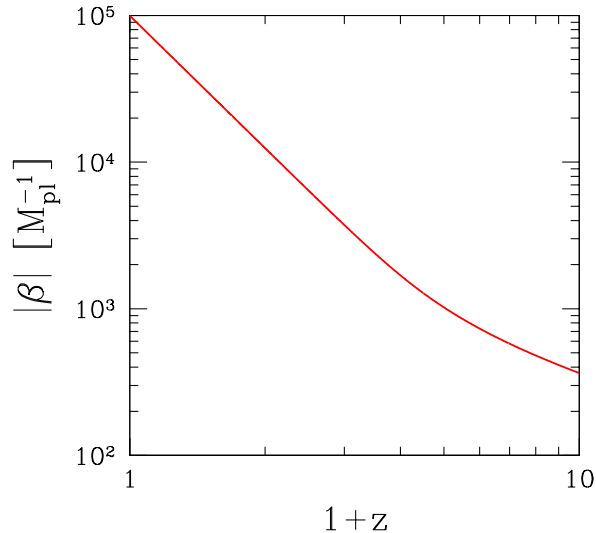}
\caption[]{The evolution of the effective coupling, $\beta$ (given
by Eq.~(\ref{eq:betalln})), as a function of redshift for the
potential Eq.~(\ref{eq:lln}). We have used
$\kappa=1\times 10^{20} M_{\rm pl}^{-1}$ and
$V_0=8.1\times10^{-13}$eV$^4$.} \label{fig:fig4.5}
\end{center}
\end{figure}

According to Eq.~(\ref{eq:masslln}) this model is characterised by a
field dependent coupling,
\begin{equation}
\label{eq:betalln}
\beta(\phi)=\frac{1}{m} \frac{\partial m}{\partial\phi}=-\frac{1}{\phi},
\end{equation}
which corresponds to a time evolution as plotted in
fig.~\ref{fig:fig4.5}.

Since the value of $\phi$ decreases with time (cf.
fig.~\ref{fig:fig3}) this means the rate of energy transfer between
the scalar field and the neutrinos and also the attraction felt
between neutrinos increases with time. Consequently, both the
neutrino mass $m_\nu$ in Eq.~(\ref{eq:masslln}) and according to
Eq.~(\ref{eq:inte}) also the neutrino energy density blow up when
$\phi$ approaches zero. Thus, from these qualitative considerations
it can already be expected that the model will run into stability
problems in the non-relativistic neutrino regime.

Secondly, we consider an inverse power-law model, which we will
refer to as the \emph{power-model}.
\begin{equation}
 \centering
 V_\phi=\frac{M^{n+4}}{\phi^n},
 \label{eq:pow}
\end{equation}
Note that this is similar to a model proposed in the context of
chameleon cosmologies~\cite{Khoury:2003aq,Brax:2004qh,shaw}.
However, there are some notable differences: Our potential does not
reduce to a cosmological constant in the asymptotic future and
$V(\phi) \to \infty$ for $\phi \to 0$. The last point is, however,
not problematic from a cosmology point of view since for realistic
value of the power-law exponent $n$ it is always true that
$\Omega_\phi \to 0$ for $t\to 0$.

\begin{figure}
\begin{center}
\includegraphics*[width=9cm]{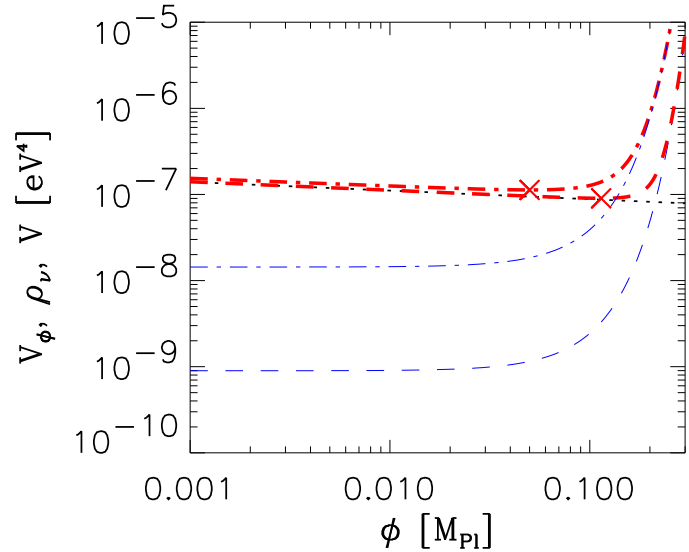}
\caption[]{The effective potential $V$ (thick lines), composed of
the scalar potential $V_\phi$ (dotted) and the neutrino energy
density $\rho_\nu$, plotted for two different redshifts, $z=1$
(dot-dashed), $z=0$ (dashed). The VEV of $\phi$ tracks the minimum
of $V$ (marked by X) and evolves to larger values for decreasing
redshift. We have used $n=0.01$ and $M=0.011$ eV.} \label{fig:fig4}
\end{center}
\end{figure}

The mass parameter $M$ is fixed by the requirements $\Omega_{DE}\sim
0.7$ and $m_\phi\gg H$. In fig.~\ref{fig:fig4} the evolution of
$\phi$ is plotted according to Eq.~(\ref{eq:eff}). In contrast to
the first model, the expectation value of $\phi$ is increasing with
time.

\begin{figure}
\begin{center}
\includegraphics*[width=8cm]{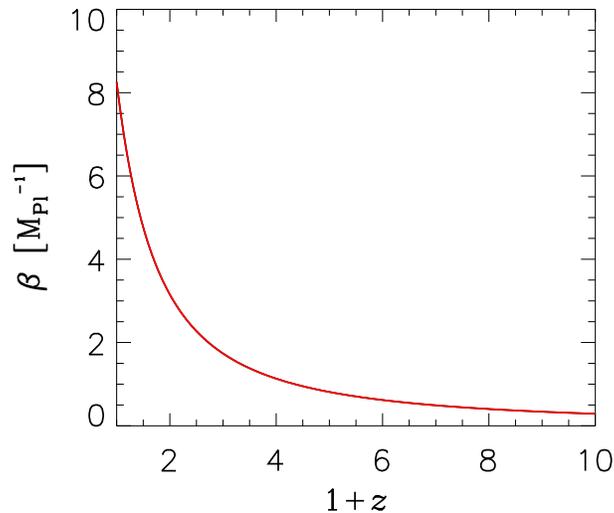}
\caption[]{The evolution of the effective coupling, $\beta$ (given
by Eq.~(\ref{eq:betapow})), as a function of redshift for the
potential Eq.~(\ref{eq:pow}). We have used $\sigma=100 M_{\rm
pl}^{-2}$.} \label{fig:fig4.6}
\end{center}
\end{figure}

In this model the dependence of the neutrino mass on the scalar field is taken to be,
\begin{equation}
 \centering
 m_\nu=m_0e^{\sigma\phi^2},
 \label{eq:masspow}
\end{equation}
 where $\sigma$ is a constant.
The power-law model is characterised by a field-dependent coupling,
\begin{equation}\label{eq:betapow}
\beta=\frac{1}{m_\nu}\frac{\partial m_\nu}{\partial \phi}={\rm
2\sigma\phi},
\end{equation}
which corresponds to a time evolution as plotted in
fig.~\ref{fig:fig4.6}.

Since according to fig.~\ref{fig:fig4} the value of $\phi$ is
increasing until the present time, the mass and consequently also
the coupling is growing with time - cf. fig.~\ref{fig:fig4.6}. The
growth of the mass depends on the choice of the parameter $\sigma$
and can be quite moderate compared to the log-linear model. This can
make the model more stable.

\section{Results} 

In this section we present the numerical results of our stability
analysis for the two MaVaN models of the last section. They are
obtained from modifying the CMBFAST code~\cite{CMBFAST} to include a
light scalar field coupled to neutrinos and were checked by altering
the CAMB code~\cite{CAMB} accordingly. We assume a neutrino energy
density of $\Omega_\nu\sim 0.02$, which corresponds roughly to the
current conservative upper limit on the sum of neutrino masses from
CMB and LSS data \cite{Spergel:2006hy,Tegmark:2006az,Goobar:2006xz}
\footnote{Note those constraints were obtained assuming
non-interacting neutrino models. Hence this assumption could be
relaxed.}, where we take the present day normalised Hubble expansion
rate to be $h=0.7$. $\Omega_\nu$ corresponds to the energy density
of three neutrino species with degenerate mass $m_{\nu_i}(z=0)\sim
0.312\,{\rm eV}\,\gg T_{\nu_0}$, which are highly non-relativistic
today. We note that this particular choice of neutrino mass has no
qualitative impact on our results.

\subsection{\label{sec:log}Log-linear Model}

The log-linear model is defined by Eq.~(\ref{eq:lln}) and
Eq.~(\ref{eq:masslln}). By fine-tuning the parameter $V_0$ for a
fixed value of $\kappa=10^{20} M_{\rm pl}^{-1}$ in Eq.~(\ref{eq:lln}),
standard cosmology with $\Omega_{\rm DE}=0.7$, $\Omega_{\rm
CDM}=0.25$, and $\Omega_{\rm b}=0.05$ at present can be
accomplished, where $\Omega_{\rm DE}=\Omega_\nu+\Omega_\phi$.

The mass of $\phi$ at present determined from Eq.~(\ref{eq:mphi}) is
$m_\phi=5.74$  ${\rm Mpc}^{-1}$$\gg H$.  Consequently, the Compton
wavelength of the scalar field, $m_\phi^{-1}$, sets the scales on
which the perturbations in (non-relativistic) MaVaNs are adiabatic,
$m_\phi^{-1}<a/k<H^{-1}$ (cf. the discussion in
sec.~\ref{sec:general}). We produce our results on a scale $k=0.1
{\rm Mpc}^{-1}\sim m_\phi$.
\begin{figure}
\begin{center}
\includegraphics*[width=10cm]{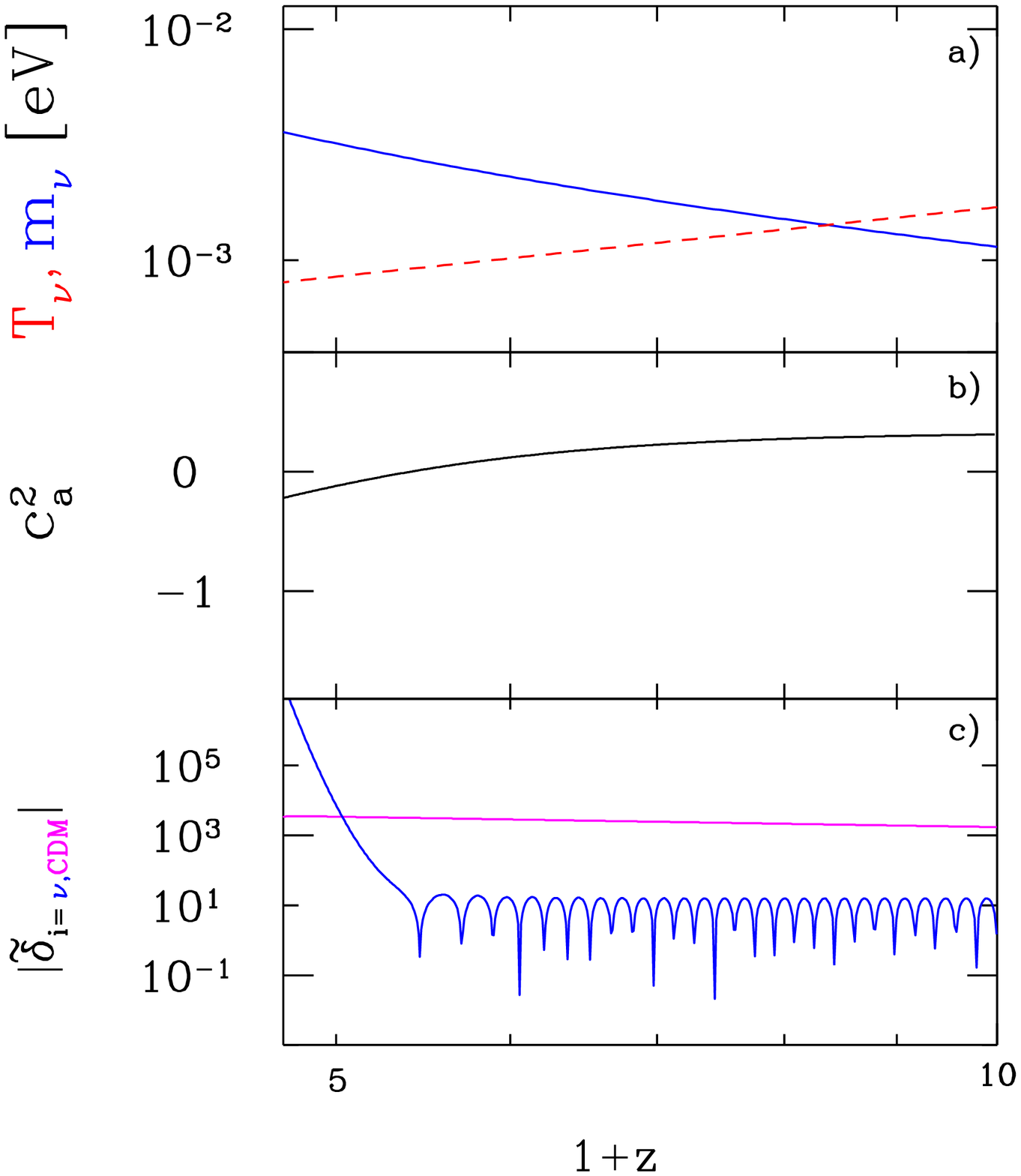}
 \caption{$a)$ Neutrino mass $m_\nu$ (solid) and temperature $T_\nu$
(dotted) as a function of redshift.  $b)$ Total dark energy sound
speed squared $c^2_a$ as a function of redshift. $c)$ Density
contrast in neutrinos (oscillating) $\delta_\nu$ and density
contrast in CDM $\delta_{\rm CDM}$ as a function of redshift on a
scale $k=0.1\,{\rm Mpc}^{-1}$. We have used $\kappa=1\times10^{20}
M_{\rm pl}^{-1}$ and $V_0=8.1\times10^{-13}$eV$^4$.}
\label{fig:Lln.eps}
\end{center}
\end{figure}
In fig.~\ref{fig:Lln.eps} we present our results for the evolution
of the neutrino mass, the sound speed squared and the density
contrast to be discussed in the following.
\begin{itemize}
\item[a)] The evolution of the neutrino mass $m_\nu(z)$ and the
 neutrino temperature $T_\nu(z)=T_{\nu_0}(1+z)$ is plotted as a
 function of redshift. As long as $m_\nu(z)\ll T_\nu(z)$, the
 neutrinos are relativistic, whereas for $m_\nu(z)\gg T_\nu(z)$ they
 have turned non-relativistic. The transition takes place at roughly
 $z+1\sim 7$, i.e.\ when $m_\nu(z) \simeq 3T_\nu(z)$. One
 interesting feature is that for $z \to 0$ the neutrino mass grows
 as $m_\nu(z) \propto a^3$ so that $\rho_\nu \to {\rm Constant}$.

\item[b)] A plot of the total adiabatic sound speed squared of the coupled
  fluid $c_a^2$. It decreases when the neutrinos approach the non-relativistic
  regime $m_\nu(z)\gg T_\nu(z)$(cf. $a)$). This is due to the drop in the
  neutrino pressure from initially $P_\nu\sim 1/3$ to $P_\nu\sim 0$ well after
  the transition of regimes.

\item[c)] A plot of the density contrast in neutrinos $\delta_\nu=\delta
  \rho_{\nu}/\rho_\nu$, and cold dark matter (CDM) $\delta_{\rm CDM}=\delta
  \rho_{\rm CDM}/\rho_{\rm CDM}$ on a scale of $k=0.1 $ Mpc$^{-1}$. As long as
  the neutrinos are still relativistic ($m_\nu(z)\ll T_\nu(z)$ cf.  $a)$), the
  perturbations in the strongly coupled scalar-neutrino fluid oscillate like
  sound waves. However, after pressure cannot offset the attractive force
  anymore ($m_\nu(z)> 3T_\nu(z)$), the neutrino density contrast blows up and
  thus grows at a much faster rate than the density contrast in CDM (the fast
  growth sets in after the effective sound speed squared has turned
  negative). This can be understood by considering the evolution of the
  coupling $\beta$ between the scalar field and neutrinos
  (cf. fig.~\ref{fig:fig4.5}), since $\beta^2$ according to
  Eq.~(\ref{denseom}) governs the evolution of the density contrast in
  non-relativistic neutrinos. The choice of a large $\kappa$ corresponds to
  $\phi \ll M_{\rm pl}$ at late times, and hence $\beta^2$ is driven to larger
  and larger values, while the VEV of $\phi$ approaches zero (cf. the
  discussion in the last section). Accordingly, $\delta_\nu$ is subject to an
  effective Newton's constant $\Geff\gg G$ (cf. the discussion in
  sec.~\ref{sec:Perturbations}). However, $\delta_{\rm CDM}$ behaves
  essentially as in General Relativity, as long as the modification to the
  gravitational effect on CDM caused by the scalar-field induced change in the
  neutrino density contrast is not prominent. Since the coupling and thus
  $\Geff$ rapidly increase with time, the scalar field transfers more and more
  energy to the neutrinos causing $m_\nu$ to increase (cf. $a)$). Therefore,
  both $\beta$ as well as the energy density in neutrinos increase such that
  the stabilising effect of the CDM becomes less and less important and
  finally becomes entirely negligible.

As a further consequence, the attraction between neutrinos also rises
steadily, while the neutrino pressure drops and ceases to stabilise the
perturbations. As demonstrated in $b)$ the total sound speed squared is thus
quickly driven to negative values, causing $\delta_\nu$ to grow faster than
exponentially (cf. also the discussion in sec.~\ref{sec:Perturbations}). As a
result, we can show that the neutrino density contrast has already turned
non-linear at $z+1\sim5$. Hence we take into account the normalisation of the
CDM density contrast which gives us a rough estimate for the normalisation of
$\delta_\nu$. As long as the dimensionless power spectrum $\Delta^2(k)=k^3
P(k)/(2\pi^2)\propto\delta_{\rm CDM}^2<1$, CDM perturbations on a scale $k$
are linear, where $P(k)$ denotes the power spectrum of CDM. Since on the
considered scale of $k=0.1\,{\rm Mpc^{-1}}$ we have $\Delta^2(k)\sim
0.3-0.4$~\cite{Percival:2006gt} for CDM, we can infer that for neutrinos
$\Delta^2(k)\propto \delta_\nu^2 >1$, when $\delta_\nu$ exceeds $\delta_{\rm
  CDM}$ by more than a factor of $\sqrt2$. This is the case at roughly
$1+z\sim 5$, while afterwards linear perturbation theory breaks down. It is
thus likely that neutrinos in this model are subject to the formation of
non-linear structure in the neutrino energy density~\cite{Afshordi:2005ym}
before the present time.
\end{itemize}

Our numerical results presented in fig.~\ref{fig:Lln.eps}
demonstrate that the total sound speed squared in the log-linear
model is negative at late times, corresponding to a fast growth of
perturbations. Thus, inevitably, the neutrino density contrast at
some point in time will go non-linear and the model becomes unstable
with the possible outcome of the formation of neutrino bound
states~\cite{Afshordi:2005ym}.

\subsection{\label{sec:Power}Power-law Potential}

\begin{figure}
\vspace* {0.0in}
\begin{center}
\includegraphics*[width=10cm]{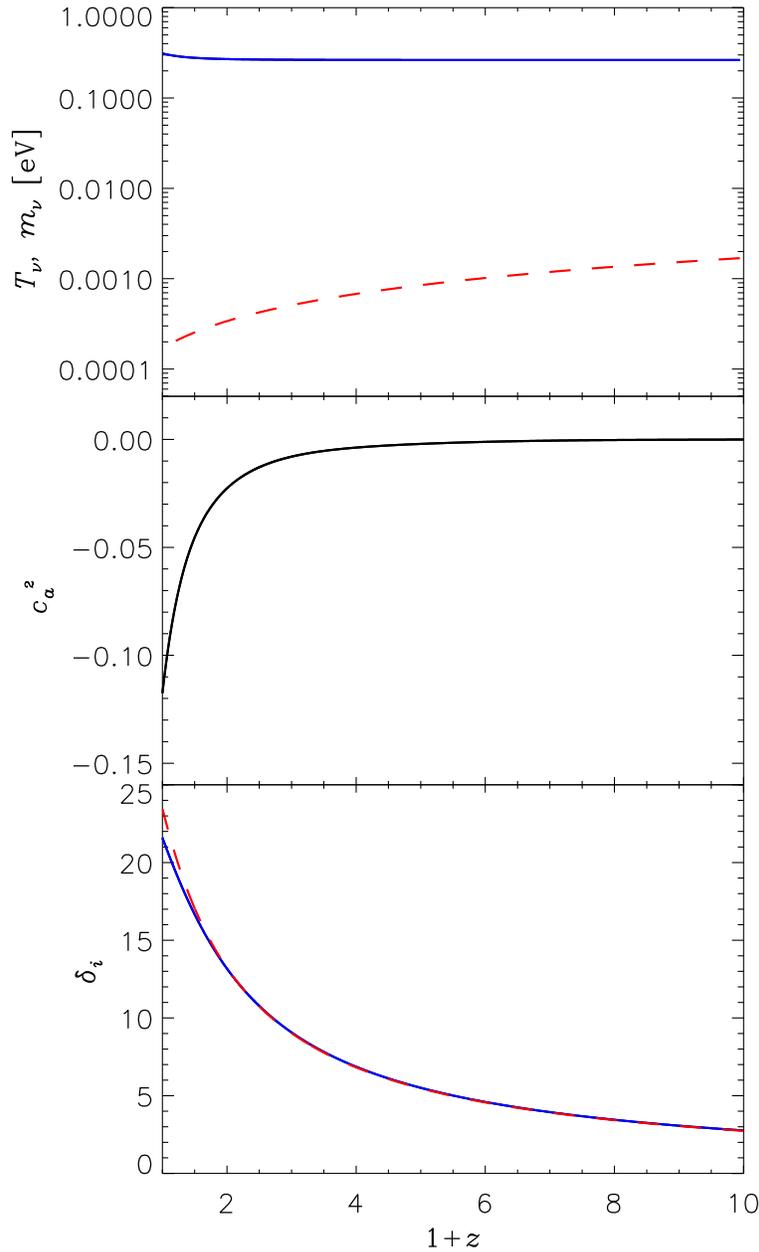}
\caption[]{$a)$ Neutrino mass $m_\nu$ (solid) and temperature
$T_\nu$ (dotted) as a function of redshift. $b)$ Total dark energy
sound speed squared $c^2_a$ as a function of redshift. $c)$ Density
contrast in neutrinos (dashed) $\delta_\nu$, and density contrast in
CDM $\delta_{\rm CDM}$ (solid) as a function of redshift on a scale
$k=10^{-3}\,{\rm Mpc}^{-1}$. We have used $n=0.01$ and $M=0.0021$
eV.} \label{fig:powerall.eps}
\end{center}
\end{figure}

\begin{figure}
\vspace* {0.0in}
\begin{center}
\includegraphics*[width=10cm]{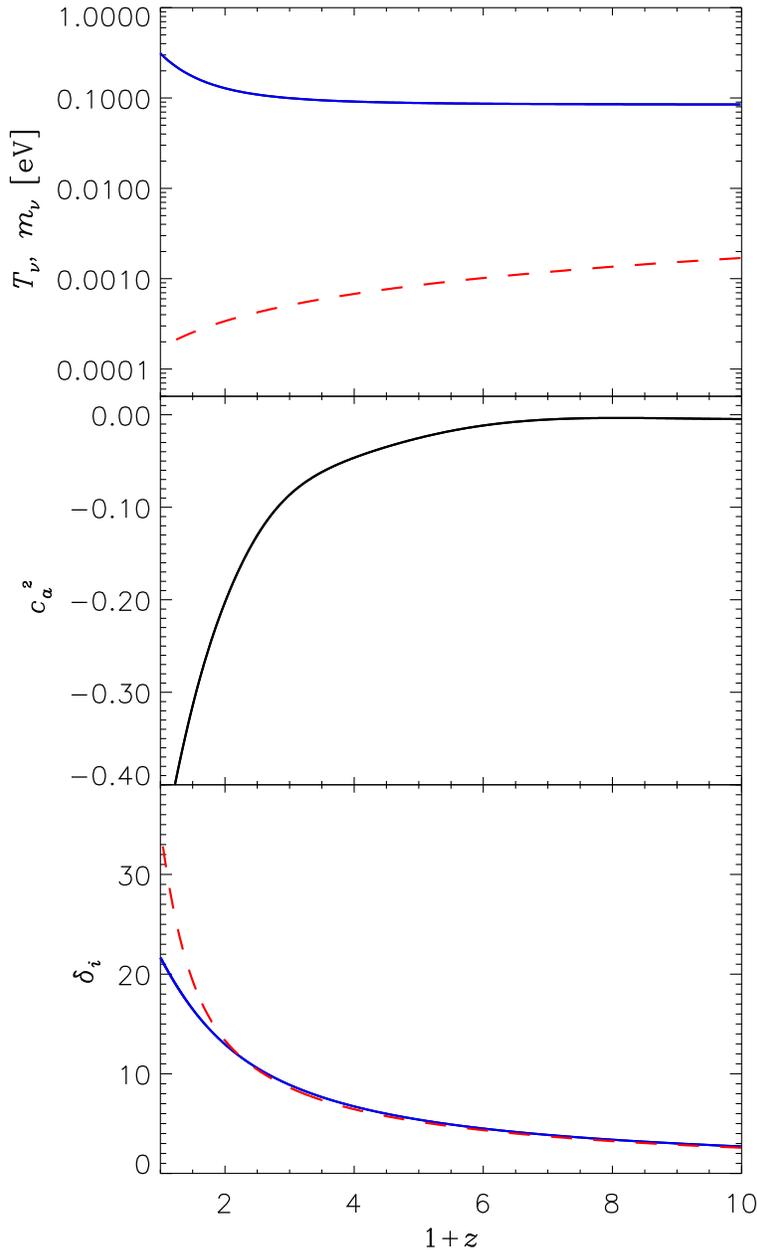}
\caption[]{$a)$ Neutrino mass $m_\nu$ (solid) and temperature
$T_\nu$ (dotted) as a function of redshift. $b)$ Total dark energy
sound speed squared $c^2_a$ as a function of redshift. $c)$ Density
contrast in neutrinos (dashed) $\delta_\nu$, and density contrast in
CDM $\delta_{\rm CDM}$ (solid) as a function of redshift on a scale
$k=10^{-3}\,{\rm Mpc}^{-1}$. We have used $n=0.1$ and $M=0.011$ eV.}
\label{fig:powerall2.eps}
\end{center}
\end{figure}

The power-law potential is defined by Eq.~(\ref{eq:pow}) and
Eq.~(\ref{eq:masspow}). We have chosen $n=0.1$ and $n=0.01$ in
Eq.~(\ref{eq:pow}) to guarantee an adiabatic evolution of the scalar
field until the present time, where $m_\phi\sim 10^{-3}{\rm
Mpc}^{-1}\sim 10H$ today. In addition, this choice of parameters
allows us to push the scales where possible adiabatic instabilities
can occur, $m_\phi^{-1}\lesssim a/k<H^{-1}$, into the linear regime.
Accordingly, we perform our perturbation analysis on a scale
$k=m_{\phi}\sim 10^{-3}\,{\rm Mpc}^{-1}$ and illustrate our results
in fig.~\ref{fig:powerall.eps} and fig.~\ref{fig:powerall2.eps} to
be described in the following.

\begin{itemize}
\item[a)] The evolution of the neutrino mass $m_\nu(z)$ and the neutrino temperature
$T_\nu(z)$ in the non-relativistic regime $m_\nu(z)\gg T_\nu(z)$ is
plotted as a function of redshift. Note that the neutrinos turn
non-relativistic at quite early times compared to the log-linear
model.

\item[b)] The evolution of the total sound speed squared $c_a^2$ of the coupled dark energy
fluid is plotted as a function of redshift. We observe that, $c_a^2$
takes negative values in the highly non-relativistic regime of the
neutrinos.

\item[c)]
The density contrast in neutrinos $\delta_\nu$, and cold dark matter
$\delta_{\rm CDM}$ is plotted on a scale of $k=10^{-3}$ Mpc$^{-1}$
for $n=0.1$ and $n=0.01$, respectively. For both cases the neutrino
mass variation is most severe at late times leading to a large
coupling at late times. In the case of $n=0.1$ the coupling is so
large at present that instabilities have effectively set in (cf. the
discussion in the log-linear model), whereas in the case of $n=0.01$
the model can be regarded as stable until the present time.

It is found that the density contrast in MaVaNs grows just as in
uncoupled neutrinos in General Relativity as long as the coupling
remains moderate. The reason is that the effects of the scalar field
on the neutrino perturbations are subdominant with respect to the
gravitational influence of CDM and baryons. As a result, the growth
of $\delta_\nu$ with time remains moderate and $\delta_{\nu}$
turns out to be of comparable size as $\delta_{\rm CDM}$
today. It has to be noted that we are looking at large scales on
which perturbations are suppressed by the large value of $k^{-1}$.
However, compared to a steeply growing coupling this effect is small
as can also be seen towards the present for the case of $n=0.1$
where the mass variation is large.

In addition, according to our analytical calculation in
sec.~\ref{sec:Perturbations} the stability condition is fulfilled on
scales were possible instabilities grow fastest. As argued in
sec.~\ref{sec:log}, the CDM perturbations are known to be linear at
the scale considered and thus the neutrino perturbations can also be
viewed as linear until the present time in the case of n=0.01.

\end{itemize}


In conclusion, fig.~\ref{fig:powerall.eps} and
fig.~\ref{fig:powerall2.eps} demonstrate that the adiabatic
power-law model is characterised by a growing mass at late times.
For moderate coupling strengths, the neutrino density contrast
follows the cold dark matter density contrast and the model can be
regarded as stable - even in the case of imaginary sound speed.
However, it depends on the choice of the model parameters whether
the model will remain stable in the future.

These results extends the analytical considerations of
Ref.~\cite{Afshordi:2005ym}. The nonlinear collapse does not happen
as soon as the neutrinos become non-relativistic, as baryons and
especially CDM, are able to attract the neutrinos in their potential
wells formed through conventional gravitational collapse. It is
important to consider the magnitude and the growth rate of the
scalar field-neutrino coupling and to compare its importance
relative to other sources of gravitational attraction. As indicated
in the previous section, the comparison can be made quantitatively
through Eq.~(\ref{denseom}).

\begin{figure}
\vspace* {0.0in}
\begin{center}
\includegraphics*[width=10cm]{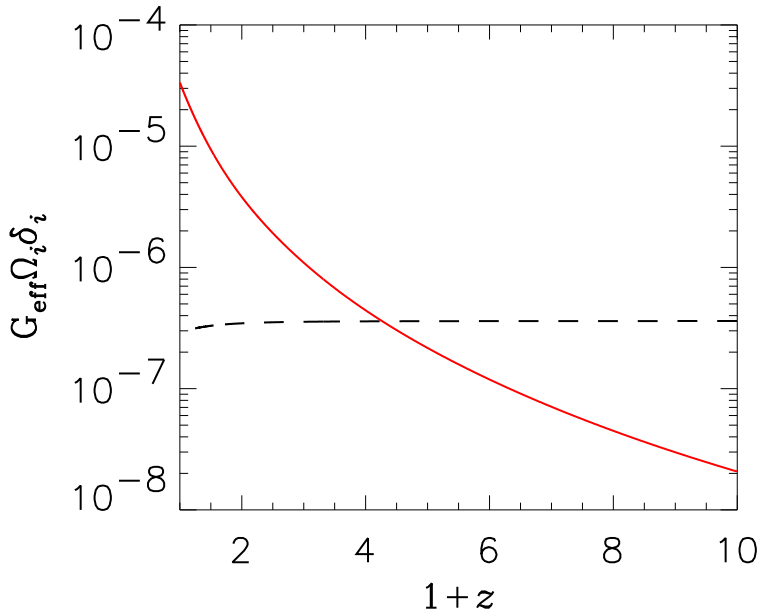}
\caption[]{Comparison of the terms from Eq.~(\ref{denseom})
$\Omega_{\nu}\frac{G_{\rm eff}}{G}\delta_\nu$(solid) and
$\Omega_{\rm CDM}\delta_{\rm CDM}$(dashed) as a function of redshift
on a scale $k=10^{-3}\,{\rm Mpc}^{-1}$. For $n=0.1$ the coupling
term is larger than the cdm term from a redshift of $z+1\sim 4$. We
have used $M=0.0021 \frac{M_{\rm pl}^{-1/2}}{{\rm Mpc}^{-1/2}}$ and
$\sigma=100 M_{\rm pl}^{-2}$.} \label{fig:compare27}
\end{center}
\end{figure}

This conclusion is further underlined by fig.\ref{fig:compare27} in
which we can see the cold dark matter term $\Omega_{\rm
CDM}\delta_{\rm CDM}$ from Eq.~(\ref{denseom}) dominating over the
coupling term $\Omega_{\nu}\frac{G_{\rm eff}}{G}\delta_\nu$ at early
times. In this regime the model is stable. At later times the
coupling term overtakes the cdm term as the coupling increases. This
effect clearly makes the $n=0.1$ model unstable.

It should be noted that in both cases for the power-law model, the
scalar field mass is decreasing such that in the very near future
$m_\phi<H$. This means that the scale on which the perturbations are
adiabatic will quickly be pushed outside the horizon and the
perturbations become non-adiabetic on all scales - for a discussion
of the stability of non-adiabatic models see \cite{Bean:2007ny}.

\subsection{A no-go theorem for mass varying neutrinos?}

In the following, we will comment on a no-go theorem in
Ref.~\cite{Afshordi:2005ym} which states that any realistic adiabatic MaVaN
model with $m_\phi^2>0$ cannot be stable for non-relativistic neutrinos.

For its deduction the authors of Ref.~\cite{Afshordi:2005ym} proceeded in the
following way. They derived an expression for the total sound speed squared
$c_a^2$ in the kinetic theory picture for $p_\nu\ll m_\nu$ assuming the
perturbations to be plane waves.  Independent of the choice of the
scalar-neutrino coupling and the scalar potential which characterise a MaVaN
model, $c_a^2$ turned out to be negative. No reference was made of the
relative gravitational importance of other relevant cosmic components like CDM
and baryons.

In the present work we have shown examples of models which demonstrate that a
detailed analysis of the potential and coupling functions as well as an
assessment of the influence of CDM and baryons, are necessary in order to
predict the growth of structure in neutrinos. In sec.~\ref{sec:Perturbations}
we found that the density contrast in neutrinos in the small scale limit only
grows exponentially if the scalar-neutrino coupling is larger than all other
relevant parameters, leading to negligible growth-slowing effects as provided
by cosmic expansion and CDM gravitational drag.

In this case we verified numerically for the log-linear model of the last
section that $c_a^2$ turns negative in agreement with the result of
\cite{Afshordi:2005ym}. We would like to point out that finite temperature
effects which can play a crucial role for the stability of a MaVaN
model~\cite{Takahashi:2006jt} were included in our calculation.

However, as demonstrated by the result for the power-model, for a moderate
coupling, the evolution of the neutrino density contrast is very similar to
the uncoupled case in ordinary General Relativity.  This was shown even in the
case of an imaginary sound speed of the dark energy fluid. Accordingly, the
perturbations were found to grow much slower than exponentially with time.

We would furthermore like to point out that our numerical analysis was very
conservative in the sense that it assumes a comparatively large neutrino mass
scale of $\sum m_{\nu}\sim 1$ eV. Thus, we would like to stress that the
stabilising effects exerted by other cosmic components on the MaVaN
perturbations can be much more efficient, in case the absolute neutrino mass
scale realized in nature turns out to be in the sub-eV range.

Based on our analysis we conclude that viable adiabatic MaVaN models
can be found which are stable until the present time. We indicate
$\Omega_{\nu}\ll \Omega_{\rm CDM}$ as the main cause, since it
enhances the stabilising influence exerted by CDM on the neutrino
density contrast. Consequently, the dynamics in stable models are
governed by CDM, largely independent of the sign of the sound speed
squared, even in the highly non-relativistic regime.

Furthermore, we have integrated the relevant equations using CMBFAST
and CAMB which work in the linear regime. Consequently, the mass of
the scalar field had to be chosen small enough (however $\gg H$) to
push the scales where possible instabilities could occur into the
linear regime.

By increasing the scalar field mass and thus reducing the range of
the scalar field, we would expect a local scalar field induced
enhancement of the gravitational clustering of neutrinos in the
non-linear regime (on scales where neutrino free-streaming cannot
inhibit the growth of perturbations). Accordingly, resulting
neutrino bound states would be interpreted as a contribution to the
CDM small scale structure, which however, on average does not affect
the equation of state of neutrino dark energy. Similarly, in
chameleon cosmologies such an enhanced small scale growth of the CDM
density contrast is predicted~\cite{green} due to the coupling to a
scalar field with range $a/k=250$ pc today. We thus refer to another
interesting class of possibly stable MaVaN models characterised by a
much larger scalar field mass. However, the detailed discussion of
these models and their phenomenological implications lies beyond the
scope of this paper.

\section{Discussion} 

Models of neutrinos coupled to a light scalar field have been
invoked to naturally explain the observed cosmic acceleration as
well as the origin of dynamical neutrino masses. However, the class
of MaVaN models characterised by an adiabatic evolution of
perturbations in the non-relativistic regime may suffer from
instabilities and as a result cease to act as dark energy. In this
paper we analysed the stability issue in the framework of linear
perturbation theory. 

For this purpose we derived the equation of
motion of the density contrast in the non-relativistic neutrino
regime in terms of the characteristic MaVaN functions, namely the
scalar potential, the scalar-neutrino coupling, and the source terms
provided by CDM and baryons. Furthermore, we modified both the
CMBFAST~\cite{CMBFAST} and CAMB~\cite{CAMB} code to include a light
scalar field coupled to neutrinos and numerically focused on two
significant MaVaN models.

We found that the instabilities in the neutrino density contrast
only occur if the influence of the scalar-neutrino coupling on the
dynamics of the perturbations dominates over the growth-slowing
effects (dragging) provided by CDM. As long as the coupling is
moderate, the neutrinos feel a gravitational drag towards the
potential wells formed by CDM. This effect can postpone the
instabilities and stabilise the perturbations until today as long as
the coupling remains of a moderate size. This result is largely
independent of the sign of the sound speed squared.

These results were obtained from considering representative limiting
cases for the time dependence of the coupling. At first, we
investigated MaVaN models characterised by a strong growth of the
coupling and thus of the neutrino masses with time. In this case, at
late times any growth-slowing effects on the perturbations provided
by the cosmic expansion or the gravitational drag of CDM can be
neglected. Consequently, independent of the choice of the scalar
potential, the analytic equation for the evolution of the neutrino
density contrast at late times involved a faster than exponentially
growing solution. Our numerical results for such a model with
logarithmic scalar potential illustrated that the onset of the
instability is around the time when the neutrinos turn
non-relativistic. In this case, the instability could be seen as the
effect of the adiabatic sound speed squared becoming negative.

Since the attraction between neutrinos increases rapidly, the sound
speed changes sign as soon as the counterbalancing pressure forces
in neutrinos have dropped sufficiently. As a result, the
non-relativistic neutrino density contrast is inevitably driven into
the non-linear regime with the likely outcome of the formation
of neutrino nuggets~\cite{Afshordi:2005ym}.

However, we demonstrated analytically that this result does not hold
true if the scalar-neutrino coupling in a MaVaN model is not strong
enough to overcompensate for the growth-slowing effects provided by
other cosmic components. More precisely, the stability condition was
found to translate into an upper bound on the scalar neutrino
coupling which is determined by the ratio between the dark matter
plus baryon density to the neutrino density.

Accordingly, the value of the allowed coupling strength depends on
the absolute neutrino mass scale realized in nature, its maximal
value being $\beta/M_{\rm pl}\sim 0(100)$ for a minimal hierarchical
neutrino mass spectrum. In this case, even though the field lies
adiabatically in the minimum of the effective potential, the
evolution of the neutrino density contrast at late times and up to
the present epoch tends to follow the CDM density contrast just as
in the uncoupled case.

Specifically, we demonstrated numerically that for the choice of a
moderately growing coupling and an inverse power law scalar
potential up to the present time the neutrino density contrast is
still in the linear regime on scales where possible instabilities
would grow fastest. Accordingly, we have identified an adiabatic
MaVaN model which can be viewed as stable until the present time.

\section*{Acknowledgments} 

LS thanks Andreas Ringwald for advice, continuous support and
discussions and Christof Wetterich and Yong-Yeon Keum for fruitful
discussions. Furthermore, we acknowledge the use of the publicly
available CMBFAST and CAMB packages~\cite{CMBFAST,CAMB}. DFM
acknowledges support from the Alexander von Humboldt Foundation and
from the Research Council of Norway through project number
159637/V30.

\section*{References} 


\begin{thebibliography}{99}


\bibitem{Bennett:2003bz}
C.~L.~Bennett {\it et al.},
Astrophys.\ J.\ Suppl.\ {\bf 148} (2003) 1

\bibitem{Spergel:2003cb}
D.~N.~Spergel {\it et al.},
Astrophys.\ J.\ Suppl.\  {\bf 148} (2003) 175

\bibitem{Spergel:2006hy}
  D.~N.~Spergel {\it et al.},
  arXiv:astro-ph/0603449.

\bibitem{Tegmark:2006az}
  M.~Tegmark {\it et al.},
  Phys.\ Rev.\  D {\bf 74}, 123507 (2006)
  [arXiv:astro-ph/0608632].

\bibitem{Riess:1998cb}
A.~G.~Riess {\it et al.}  [Supernova Search Team Collaboration],
Astron.\ J.\  {\bf 116}, 1009 (1998)



\bibitem{Perlmutter:1998np}
S.~Perlmutter {\it et al.} [Supernova Cosmology Project
Collaboration],
Astrophys.\ J.\  {\bf 517} (1999) 565

\bibitem{Astier:2005qq}
  P.~Astier {\it et al.}  [The SNLS Collaboration],
  Astron.\ Astrophys.\  {\bf 447}, 31 (2006)

\bibitem{Wood-Vasey:2007jb}
  W.~M.~Wood-Vasey {\it et al.},
  arXiv:astro-ph/0701041.

\bibitem{Wetterich:1987fm}
C.~Wetterich,
Nucl.\ Phys.\ B {\bf 302}, 668 (1988).

\bibitem{Peebles:1987ek}
P.~J.~E.~Peebles and B.~Ratra,
Astrophys.\ J.\  {\bf 325}, L17 (1988).

\bibitem{Ratra:1987rm}
B.~Ratra and P.~J.~E.~Peebles,
Phys.\ Rev.\ D {\bf 37}, 3406 (1988).

\bibitem{Zlatev:1998tr}
I.~Zlatev, L.~M.~Wang and P.~J.~Steinhardt,
Phys.\ Rev.\ Lett.\  {\bf 82}, 896 (1999)

\bibitem{Wang:1999fa}
L.~M.~Wang, R.~R.~Caldwell, J.~P.~Ostriker and P.~J.~Steinhardt,
Astrophys.\ J.\  {\bf 530}, 17 (2000)

\bibitem{Steinhardt:1999nw}
P.~J.~Steinhardt, L.~M.~Wang and I.~Zlatev,
Phys.\ Rev.\ D {\bf 59}, 123504 (1999)

\bibitem{Barreiro:1999zs}
T.~Barreiro, E.~J.~Copeland and N.~J.~Nunes,
Phys.\ Rev.\ D {\bf 61}, 127301 (2000)

\bibitem{Baccigalupi:2001aa}
C.~Baccigalupi, A.~Balbi, S.~Matarrese et al.
Phys.\ Rev.\ D {\bf 65}, 063520 (2002)

\bibitem{Caldwell:2003vp}
R.Caldwell, M.Doran, C.Mueller, G.Schaefer and
C.Wetterich,
AJ.{\bf 591},L75(2003)

\bibitem{Mota:2004pa}
  D.~F.~Mota and C.~van de Bruck,
  Astron.\ Astrophys.\  {\bf 421} (2004) 71

\bibitem{Amendola:1999er}
L.~Amendola,
Phys.\ Rev.\ D {\bf 62}, 043511 (2000)



\bibitem{Bertolami:1999dp}
O.~Bertolami and P.~J.~Martins,
Phys.\ Rev.\ D {\bf 61}, 064007 (2000)

\bibitem{tomi}
  T.~Koivisto and D.~F.~Mota,
  Phys.\ Lett.\  B {\bf 644} (2007) 104
  arXiv:astro-ph/0606078; ibidem Phys.\ Rev.\  D {\bf 75}, 023518 (2007).

\bibitem{Perrotta:1999am}
F.~Perrotta, C.~Baccigalupi and S.~Matarrese,
Phys.\ Rev.\ D {\bf 61}, 023507 (2000)

\bibitem{domenico:2002}
   L.~Amendola and D.~Tocchini-Valentini
   Phys.\ Rev.\ D {\bf 66}, 043528 (2002)

\bibitem{domenico}
D.~Tocchini-Valentini and L.~Amendola,
  Phys.\ Rev.\ D {\bf 65} (2002) 063508


\bibitem{Armendariz-Picon:1999rj}
C.~Armendariz-Picon, T.~Damour and V.~Mukhanov,
Phys.\ Lett.\ B {\bf 458}, 209 (1999)

\bibitem{Chiba:1999ka}
T.~Chiba, T.~Okabe and M.~Yamaguchi,
Phys.\ Rev.\ D {\bf 62}, 023511 (2000)

\bibitem{Armendariz-Picon:2000ah}
C.~Armendariz-Picon, V.~Mukhanov and P.~J.~Steinhardt,
Phys.\ Rev.\ D {\bf 63}, 103510 (2001)


\bibitem{Caldwell:1999ew}
R.~R.~Caldwell,
Phys.\ Lett.\ B {\bf 545}, 23 (2002)

\bibitem{Schulz:2001yx}
A.~E.~Schulz and M.~J.~White,
Phys.\ Rev.\ D {\bf 64}, 043514 (2001)

\bibitem{Carroll:2003st}
S.~M.~Carroll, M.~Hoffman and M.~Trodden,
Phys.\ Rev.\ D {\bf 68}, 023509 (2003)




\bibitem{Deffayet:2001pu}
C.~Deffayet, G.~R.~Dvali and G.~Gabadadze,
Phys.\ Rev.\ D {\bf 65}, 044023 (2002)

\bibitem{Dvali:2003rk}
G.~Dvali and M.~S.~Turner,
arXiv:astro-ph/0301510.

\bibitem{sean}
  S.~M.~Carroll, V.~Duvvuri, M.~Trodden and M.~S.~Turner,
  Phys.\ Rev.\ D {\bf 70} (2004) 043528

\bibitem{brook}
  A.~W.~Brookfield, C.~van de Bruck and L.~M.~H.~Hall,
  Phys.\ Rev.\ D {\bf 74} (2006) 064028

\bibitem{morad}
  M.~Amarzguioui, O.~Elgaroy, D.~F.~Mota and T.~Multamaki,
  Astron.\ Astrophys.\  {\bf 454} (2006) 707

\bibitem{Hung:2000yg}
  P.~Q.~Hung,
  arXiv:hep-ph/0010126.

\bibitem{Gu:2003er}
  P.~Gu, X.~Wang and X.~Zhang,
  Phys.\ Rev.\ D {\bf 68}, 087301 (2003)


\bibitem{Fardon:2003eh}
  R.~Fardon, A.~E.~Nelson and N.~Weiner,
  JCAP {\bf 0410} (2004) 005

\bibitem{Peccei:2004sz}
  R.~D.~Peccei,
  Phys.\ Rev.\ D {\bf 71} (2005) 023527

\bibitem{Goobar:2006xz}
  A.~Goobar, S.~Hannestad, E.~Mortsell and H.~Tu,
  JCAP {\bf 0606}, 019 (2006)

\bibitem{Elgaroy:2006iy}
  O.~Elgaroy,
  arXiv:hep-ph/0612097.

\bibitem{skordis}
 C.~Skordis, D.~F.~Mota, P.~G.~Ferreira and C.~Boehm,
  Phys.\ Rev.\ Lett.\  {\bf 96}, 011301 (2006).

\bibitem{Zunckel:2006mt}
  C.~Zunckel and P.~G.~Ferreira,
  arXiv:astro-ph/0610597.


\bibitem{Fogli:2006yq}
  G.~L.~Fogli {\it et al.},
  arXiv:hep-ph/0608060.

\bibitem{Hannestad:2006mi}
  S.~Hannestad and G.~G.~Raffelt,
  JCAP {\bf 0611}, 016 (2006)

\bibitem{Feng:2006zj}
  B.~Feng, J.~Q.~Xia, J.~Yokoyama, X.~Zhang and G.~B.~Zhao,
  arXiv:astro-ph/0605742.

\bibitem{Seljak:2006bg}
  U.~Seljak, A.~Slosar and P.~McDonald,
  JCAP {\bf 0610}, 014 (2006)

\bibitem{Lesgourgues:2006nd}
  J.~Lesgourgues and S.~Pastor,
  Phys.\ Rept.\  {\bf 429}, 307 (2006)


\bibitem{Hannestad:2006zg}
  S.~Hannestad,
  arXiv:hep-ph/0602058.

\bibitem{Hannestad:2005gj}
  S.~Hannestad,
  Phys.\ Rev.\ Lett.\  {\bf 95}, 221301 (2005)

\bibitem{Hannestad:2003xv}
  S.~Hannestad,
  JCAP {\bf 0305}, 004 (2003)

\bibitem{Khoury:2003aq}
  J.~Khoury and A.~Weltman,
  Phys.\ Rev.\ Lett.\  {\bf 93} (2004) 171104


\bibitem{Brax:2004qh}
  P.~Brax, C.~van de Bruck, A.~C.~Davis, J.~Khoury and A.~Weltman,
  Phys.\ Rev.\ D {\bf 70} (2004) 123518

\bibitem{shaw2}  D.~F.~Mota and D.~J.~Shaw,
Phys.\ Rev.\  D {\bf 75}, 063501 (2007).
[arXiv:hep-ph/0608078].


\bibitem{Schrempp:2006mk}
  L.~Schrempp,
  arXiv:astro-ph/0611912.

\bibitem{Ringwald:2006ks}
  A.~Ringwald and L.~Schrempp,
  JCAP {\bf 0610}, 012 (2006)

\bibitem{Kaplan:2004dq}
  D.~B.~Kaplan, A.~E.~Nelson and N.~Weiner,
  Phys.\ Rev.\ Lett.\  {\bf 93}, 091801 (2004)

\bibitem{Barger:2005mn}
  V.~Barger, P.~Huber and D.~Marfatia,
  Phys.\ Rev.\ Lett.\  {\bf 95}, 211802 (2005)

\bibitem{Cirelli:2005sg}
  M.~Cirelli, M.~C.~Gonzalez-Garcia and C.~Pena-Garay,
  Nucl.\ Phys.\ B {\bf 719}, 219 (2005)

\bibitem{Barger:2005mh}
  V.~Barger, D.~Marfatia and K.~Whisnant,
  Phys.\ Rev.\ D {\bf 73}, 013005 (2006)

\bibitem{Gu:2005pq}
  P.~H.~Gu, X.~J.~Bi, B.~Feng, B.~L.~Young and X.~Zhang,
  arXiv:hep-ph/0512076.

\bibitem{Li:2004tq}
  H.~Li, Z.~g.~Dai and X.~m.~Zhang,
  Phys.\ Rev.\ D {\bf 71}, 113003 (2005)


\bibitem{Afshordi:2005ym}
  N.~Afshordi, M.~Zaldarriaga and K.~Kohri,
  Phys.\ Rev.\ D {\bf 72} (2005) 065024

\bibitem{Brookfield:2005td}
A.~W.~Brookfield, C.~van de Bruck, D.~F.~Mota and
D.~Tocchini-Valentini,
Phys.\ Rev.\ Lett.\  {\bf 96} (2006) 061301

\bibitem{Brookfield:2005bz}
A.~W.~Brookfield, C.~van de Bruck, D.~F.~Mota and
D.~Tocchini-Valentini,
Phys.\ Rev.\ D {\bf 73} (2006) 083515

\bibitem{Tonry:2003zg}
  J.~L.~Tonry {\it et al.}  [Supernova Search Team Collaboration],
  Astrophys.\ J.\  {\bf 594} (2003) 1

\bibitem{Bean:2003fb}
  R.~Bean and O.~Dore,
  Phys.\ Rev.\ D {\bf 69}, 083503 (2004)

\bibitem{Hannestad:2005ak}
  S.~Hannestad,
  Phys.\ Rev.\ D {\bf 71}, 103519 (2005)

\bibitem{Kaplinghat:2006jk}
  M.~Kaplinghat and A.~Rajaraman,
  arXiv:astro-ph/0601517.


\bibitem{pedro}
  L.~M.~G.~Beca and P.~P.~Avelino,
  Mon.\ Not.\ Roy.\ Astron.\ Soc.\  {\bf 376} (2007) 1169.

\bibitem{Takahashi:2006jt}
  R.~Takahashi and M.~Tanimoto,
  JHEP {\bf 0605} (2006) 021

\bibitem{Spitzer:2006hm}
  C.~Spitzer,
  arXiv:astro-ph/0606034.

\bibitem{Fardon:2005wc}
  R.~Fardon, A.~E.~Nelson and N.~Weiner,
  JHEP {\bf 0603}, 042 (2006)
  [arXiv:hep-ph/0507235].

\bibitem{mb}C.-P.~Ma and E.~Bertschinger, Astrohys.\ J.\ {\bf 455}, 7
(1995).

\bibitem{Keum}
Yong-Yeon Keum, Talk at 2006 International
Symposium on Cosmology and Particle  Astrophysics, November 15-17, 2006, NTU,
Taipei, Taiwan

\bibitem{Koivisto:2005nr}
  T.~Koivisto,
  Phys.\ Rev.\ D {\bf 72} (2005) 043516


\bibitem{Amendola:2003wa}
  L.~Amendola,
  Phys.\ Rev.\ D {\bf 69}, 103524 (2004)

\bibitem{Coleman:1973jx}
  S.~R.~Coleman and E.~Weinberg,
  Phys.\ Rev.\ D {\bf 7}, 1888 (1973).

\bibitem{Kamke}
  E.~Kamke,
  Teubner Verlag, 2002.

\bibitem{shaw}
D.~F.~Mota and D.~J.~Shaw,
Phys.\ Rev.\ Lett.\  {\bf 97} (2006) 151102

\bibitem{kraus}C.~Kraus {\it et al.} European Physical Journal C
(2003), proceedings of the EPS 2003

\bibitem{CMBFAST}
U.~Seljak and M.~Zaldarriaga,
Astrophys.\ J.\  {\bf 469} (1996) 437.

\bibitem{CAMB}
  A.~Lewis and S.~Bridle,
  Phys.\ Rev.\  D {\bf 66}, 103511 (2002)
  [arXiv:astro-ph/0205436].

\bibitem{Percival:2006gt}
  W.~J.~Percival {\it et al.},
  arXiv:astro-ph/0608636.

\bibitem{green} P.~Brax, C.~van de Bruck, A.~C.~Davis and A.~M.~Green,
  Phys.\ Lett.\  B {\bf 633}, 441 (2006)

\bibitem{Bean:2007ny}
  R.~Bean, E.~E.~Flanagan and M.~Trodden,
  arXiv:0709.1128 [astro-ph].

\end{thebibliography}
\end{document}